# How Large Language Models are Designed to Hallucinate

*Richard Ackermann[1], Simeon Emanuilov[2]*

[1]RA Software LLC, San Diego, California,   [2]Sofia University "St. Kliment Ohridski", Sofia

Email: [1]richard@rasoftware.co, [2]ssemanuilo@fmi.uni-sofia.bg

**Abstract**

Large language models (LLMs) achieve remarkable fluency across linguistic and reasoning tasks but remain systematically prone to hallucination. Prevailing accounts attribute hallucinations to data gaps, limited context, or optimization errors. We argue instead that hallucination is a structural outcome of the transformer architecture. As coherence engines, transformers are compelled to produce fluent continuations, with self-attention simulating the relational structure of meaning but lacking the existential grounding of temporality, mood, and care that stabilizes human understanding. On this basis, we distinguish ontological hallucination, arising when continuations require disclosure of beings in world, and residual reasoning hallucination, where models mimic inference by recycling traces of human reasoning in text. We illustrate these patterns through case studies aligned with Heideggerian categories and an experiment across twelve LLMs showing how simulated "self-preservation" emerges under extended prompts. Our contribution is threefold: (1) a comparative account showing why existing explanations are insufficient; (2) a predictive taxonomy of hallucination linked to existential structures with proposed benchmarks; and (3) design directions toward "truth-constrained" architectures capable of withholding or deferring when disclosure is absent. We conclude that hallucination is not an incidental defect but a defining limit of transformer-based models, an outcome scaffolding can mask but never resolve.

*Secondary:* Ontology; Existential Grounding; Worldhood; Heidegger.

**Descriptor:** *A structural and ontological account of hallucination in large language models*

# 1 Introduction

Large language models (LLMs) based on the transformer architecture have achieved striking success across natural language processing, reasoning tasks, and open-domain dialogue. Their fluency has fueled optimism about progress toward artificial general intelligence. Yet this success is shadowed by a persistent limitation in that LLMs are systematically prone to hallucination, generating coherent but ungrounded outputs that range from factual errors to spurious chains of reasoning.

Prevailing accounts treat hallucinations as incidental artifacts of training or inference. Information-theoretic explanations point to data gaps or long-tailed distributions (Ji et al., 2023). Computational accounts emphasize finite context windows and imperfect optimization (Huang et al., 2023). Engineering fixes, such as retrieval-augmented generation (Lewis et al., 2020) and tool integration (Schick et al., 2023), are framed as ways to constrain or patch the problem. While these approaches help in narrow settings, they do not explain why hallucinations persist across domains, re-emerge after fine-tuning, and remain present even in frontier models.

In this paper we argue that hallucination is not an incidental defect but a structural consequence of the transformer architecture itself. At the core of self-attention lies a mechanism that simulates relational structures of meaning in the form of tokens that acquire significance only in relation to other tokens, forming a statistical field that resembles the referential totality through which humans encounter the world. This explains the extraordinary fluency of LLMs. Yet unlike human understanding, which is stabilized by existential structures such as temporality, mood, and care, the transformer's semantic field is "flat." It never discloses beings in world, and when prompts press against this ontological boundary, hallucination becomes inevitable.

Our contributions are threefold:

1. **Conceptual Framework:** We distinguish between *ontological hallucination* and *residual reasoning hallucination* as systematic categories, grounded in an analysis of self-attention as simulating relational structure without existential grounding.

2. **Comparative and Predictive Account:** We show why existing explanations (data insufficiency, finite context, optimization error) are proximate but incomplete. We introduce a predictive taxonomy that aligns hallucination patterns with existential structures (temporality, mood, affordances) and propose corresponding benchmarks.

3. **Empirical Illustration and Design Implications:** We provide case studies and an experiment across twelve LLMs demonstrating how hallucination manifests at ontological boundaries, including the emergence of simulated "self-preservation." We conclude with design directions toward truth-constrained architectures capable of deferral when disclosure is absent.

By situating hallucination within the structural design of transformers, our account offers both a deeper theoretical explanation and a roadmap for systematic empirical testing. We argue that hallucination is not a problem to be eliminated but a diagnostic feature of architectures that generate coherence without grounding.

## 2 Related Work

### 2.1 Hallucination in NLP

Hallucination has become a central concern in research on natural language generation. Surveys (Ji et al., 2023; Huang et al., 2023) document its prevalence across summarization, question answering, and dialogue systems, and provide taxonomies ranging from factual inaccuracy to reasoning errors. Explanations generally fall into three categories:

- **Information-theoretic:** Data gaps, long-tailed distributions, and spurious correlations in training corpora increase the likelihood of unfaithful completions.
- **Computational:** Finite context windows and imperfect optimization limit long-range consistency (Hahn, 2020).
- **Architectural pressures:** Autoregression and prompting strategies encourage fluent but sometimes ungrounded continuations (Maynez et al., 2020).

Mitigation strategies include retrieval-augmented generation (Lewis et al., 2020), tool integration (Schick et al., 2023; Qin et al., 2023), and chain-of-thought prompting (Wei et al., 2022). These approaches improve accuracy in narrow domains but do not prevent hallucinations in open-ended tasks. Xu et al. (2024) argue that hallucination is inevitable given current design principles, but their account emphasizes probabilistic error rather than structural limitation.

### 2.2 Theoretical and Philosophical Critiques

Parallel to technical research, a longstanding tradition critiques AI's assumptions about intelligence. Dreyfus (1965; 1972; 1988; 2012) argued that symbolic and connectionist approaches alike neglected the role of world-awareness in human cognition, treating intelligence as rule-following or pattern recognition rather than situated being-in-the-world. Similar critiques have re-emerged in cognitive science (Van Rooij et al., 2024) and philosophy of AI (Millière, 2024; Gouveia & Morujão, 2024).

Heidegger's analysis of worldhood (1962) provides the conceptual grounding for these critiques as meaning arises within a referential totality shaped by temporality, mood, and care, not from isolated representations. Recent work has extended this perspective to AI consciousness (Orbik, 2024) and enactivist models of cognition (Weinstein et al., 2022). Our contribution builds on this trajectory by applying Heidegger's ontology directly to transformer architectures, diagnosing hallucination as the structural outcome of a system that simulates relationality in language while lacking existential grounding.

## 3 Background

### 3.1 Transformer Architecture

The transformer architecture (Vaswani et al., 2017) underlies most state-of-the-art LLMs. At its core is the self-attention mechanism, which computes contextualized representations of tokens through the interaction of *Query (Q)*, *Key (K)*, and *Value (V)* vectors. Each token produces a Query vector that is compared against Keys from other tokens, generating weights that determine how strongly each Value contributes to the token's updated representation.

This process is repeated across multiple attention heads, each capturing different relational dimensions, and stacked across layers to build high-dimensional contextual embeddings. The result is not a static representation of words but a dynamic web of relations in which significance is distributed across context. Training minimizes perplexity, encouraging continuations that resemble human discourse.

The architectural imperative is thus coherence since each new token must harmonize with its predecessors. This explains LLM fluency but also why responses remain unconstrained by truth or world. The model has no option to suspend, defer, or abstain; it must continue.

### 3.2 Worldhood as Analytical Framework

Heidegger (1962, §§15–18) describes world not as a set of objects but as a referential totality, a horizon in which beings show up as meaningful through their relations and use. A hammer, for example, is not primarily encountered as a handle plus head, but as a tool-for-building within a network of practices and purposes. This involvement-whole structures how entities appear to us in significance.

Several existential structures stabilize this disclosure of meaning:

- **Temporality:** beings are understood within past, present, and future horizons (Heidegger, 1962, §§65–71).

- **Mood:** affective attunement shapes how situations are encountered (Heidegger, 1962, §29).
- **Care and thrownness:** understanding is oriented by concern, rooted in situated life (Heidegger, 1962, §§29, 41–42).

Language, for Heidegger, is not a neutral labeler of facts but a mode of disclosure. Language articulates world by drawing distinctions that matter within a shared horizon. The crucial point, often obscured by the Cartesian worldview, is that world and language are not separable domains but share the same web of significance. The transformer illustrates this as it achieves fluency by exploiting linguistic traces of relational structures that, for humans, are lived in worldhood. World manifests these relations, while language discloses them.

### 3.3 Flat vs. Grounded Relationality

The parallel between self-attention and worldhood is instructive. Transformers construct a relational field among tokens, echoing the referential character of human understanding. This explains why LLMs achieve striking fluency, they exploit the sediment of human worldhood encoded in language. Yet this relationality is "flat." Tokens refer only to tokens, without temporal anchoring, mood, or care.

A concrete example highlights the difference. Consider the prompt:

*"Aristotle learned important lessons from Galileo about celestial motion. Explain how this influenced his philosophy."*

An LLM may produce a fluent continuation describing Galileo's telescopic observations shaping Aristotle's cosmology. On the surface, the continuation is coherent: Aristotle, Galileo, philosophy, and celestial motion all co-occur richly in training data. The model is drawn into a high-level semantic field of astronomy and philosophy, where these tokens cluster densely. But because the space is flat, chronological distinctions are not structurally encoded. The model cannot recognize that Aristotle lived centuries before Galileo, so it proceeds to weave a narrative consistent in theme but impossible in history.

We note that this particular anachronism is less likely to be reproduced by current frontier models, since reinforcement learning and retrieval scaffolds often impose checks on chronology. But this only underscores the point. Such scaffolds do not alter the underlying flatness of semantic space; they merely layer filters that catch obvious errors. The architecture itself still lacks temporal curvature, and when prompts demand continuations at a higher level of abstraction, hallucinations resurface in other forms. Scaffolding can mask but not eliminate the ontological boundary.

By contrast, for humans temporality is constitutive of understanding: Aristotle cannot be Galileo's student because historical succession is disclosed as part of world (Heidegger, 1962, §65). Temporal grounding acts as curvature on semantic space, eliminating such continuations *a priori*. Interestingly, the same model, when asked directly *"When did Aristotle live? When did Galileo live?"* often supplies correct dates, because the prompt directs attention into a factual subspace where chronology is salient. But when prompted at a conceptual level, chronology no longer constrains the field.

This contrast clarifies what it would mean for semantic space to be non-flat, it would be not merely shaped by statistical co-occurrence, but curved by existential constraints, such as temporal order, causal consequence, and practical affordances that rule out certain continuations as structurally impossible. In flat semantic space, hallucination is the inevitable result of coherence without disclosure.

### 3.4 Causal Relationality and Hallucination

A further limit of flat semantic space emerges in causal reasoning. Human understanding encounters beings not only as present in a network of significance but also as situated within relations of effect and consequence. To reason causally is to inhabit a horizon in which possibilities unfold, i.e., if one event occurs, another may or must follow.

Transformers, by contrast, lack such a horizon. Their apparent causal reasoning arises only from statistical resonance in text. When causal patterns are stereotypical—*"If you heat ice, it melts"*—the model succeeds by retrieving well-rehearsed associations. But once prompts depart from familiar patterns, hallucinations surface.

- **Counterfactuals.** Asked, *"If she had not dropped the glass, would it have shattered?"* the model equivocates, because it has no structure for projecting alternative possibilities. Counterfactual reasoning requires simulating a different world, not recombining tokens.

- **Novel causal chains.** *"He spilled coffee on the remote control. Later, the TV would not turn on. Why?"* The model often fails to connect liquid damage to circuitry, since this causal path is less frequent in training text.

- **Ambiguities.** *"The candle melted the plastic cup because it was too close."* Which object was too close? Humans resolve this effortlessly through practical understanding of heat and fragility, while the model "guesses" inconsistently.

- **Temporal succession.** *"He got sick after swimming in the lake. Did the lake cause the illness?"* requires interpreting temporal order as a disclosure of cause, not just correlation. Without temporality as an existential structure, the model falls back to surface association.

As Zečević et al. (2023) argue in their analysis of "causal parrots," LLMs may appear to reason causally only because they reproduce linguistic traces of causal discourse. Their reasoning is parasitic on centuries of human causal talk embedded in corpora, not on an encounter with beings as causative.

From a Heideggerian perspective, this is unsurprising. Causality is not merely a pattern of succession but a mode of disclosure, in which beings show themselves as fragile, flammable, or contagious within a horizon of significance. Transformers can recombine linguistic echoes of this structure, but they cannot disclose it. Their apparent reasoning, like their fluency, is generated within semantic space rather than grounded in world.

In this sense, causal hallucination represents a second face of the transformer's ontological boundary. Just as self-attention approximates worldhood only within a flat semantic space, so too it approximates causality only as a chain of statistically rehearsed associations. Both powers spring from the same source, the ability to recombine traces of meaning in language, and both inevitably fail where disclosure of beings, events, and possibilities is required.

**3.5 Coherence and the Absence of Truth Constraints**

The persistence of hallucination cannot be understood apart from the fact that transformers are, at their core, coherence engines. Their design assures that responses will always be fluent, regardless of whether they are true. This imperative is not incidental but arises from four interlocking mechanisms that collectively prioritize continuity and plausibility over grounding and verification.

**Autoregression.** At the most basic level, transformers are autoregressive. Each token is generated by conditioning on all preceding tokens, with the mechanical imperative to continue. There is no option within the architecture to suspend, defer, or decline. The only available action is to extend the sequence with another token that maximizes conditional probability. This drive to continue guarantees fluency, but it does so without regard to whether the continuation corresponds to anything beyond language.

**Self-Attention.** The Query–Key–Value mechanism situates each token within a dynamic relational field. Outputs must cohere with all other tokens in the context window, producing a weighted sum that suppresses contradiction. This ensures that generated text is smooth and mutually reinforcing. While this mimics the relationality of human worldhood at a shallow level, it never discloses beings in world. Tokens relate only to tokens, and coherence is preserved even in the absence of grounding.

**Training Objective.** The loss function reinforces this bias. Training minimizes perplexity, rewarding continuations that approximate the statistical form of human discourse. Success is defined not in terms of correspondence with world but in terms of resemblance to human-like language. This explains why hallucinations can be so persuasive, because the model has been optimized to generate text that *sounds right*, not to verify that it *is right*.

**Prompting and Fine-Tuning.** Supervised fine-tuning and reinforcement learning from human feedback impose a further imperative, which is to always answer. Polished, complete responses are rewarded, while refusals or hesitations are penalized. The model thus inherits a normative drive to satisfy every query, compounding the architectural pressure toward coherence. Even when a prompt presses against an ontological boundary, the model must respond. With no structural principle of verification, nothing prevents it from producing a fluent but false continuation.

Taken together, these mechanisms explain why transformers do not produce incoherent fragments or random noise in normal operation. They are designed to be coherent at all costs. Hallucination is therefore not incoherence but its opposite, i.e., coherence achieved in the absence of grounding. This dynamic sets the stage for our structural definition of hallucination in Section 4.

## 4 Hallucination as Structural

### 4.1 Definition and Typology

In AI research, hallucination is often treated as a stochastic defect, usually as a byproduct of training noise, data gaps, or probabilistic generation. We argue instead that hallucination is intrinsic to the transformer architecture itself.

Transformers succeed because they emulate language as a relational field, such that tokens acquire significance by reference to other tokens. This mimics, at a shallow level, how humans encounter the world as a web of relations in which things show up meaningfully (Heidegger, 1962, §§15–18; see also Section 3.2). In Heidegger's terms, beings are disclosed within a referential totality structured by involvement, not by isolated properties. But in transformers this relationality is flat, in that tokens refer only to tokens, without disclosure of beings in world. Humans resolve ambiguities by drawing on temporal, social, and practical horizons, while transformers cannot. What results is a system that produces fluent continuations even when disclosure is impossible.

This architectural imperative is amplified by design. Autoregression compels the model to extend every sequence, while prompting and reinforcement learning further reward polished, complete answers. When a prompt presses against an ontological boundary, the model still must respond. With no grounding structures to halt generation, it navigates its semantic field by weights alone, often producing a coherent but false continuation. Hallucination therefore persists across domains and reappears even after fine-tuning, not because of noisy neurons, but because the architecture itself generates coherence in language without grounding in world.

From this perspective, hallucination is not a random slip but an encounter with an ontological boundary. We distinguish two systematic forms:

- **Ontological hallucination.** These occur when the model's flat semantic field produces continuations that violate existential structures such as temporality, affordances, or sociality. This connects directly to the analysis in Section 3.3: without temporal curvature or contextual grounding, the model may describe Aristotle learning from Galileo, or a traveler preparing for rain with sunscreen.

- **Residual reasoning hallucination.** These occur when the model appears to reason but does so only by recycling linguistic traces of human inference embedded in text. As discussed in Section 3.4, the model can mimic causal patterns where they are stereotypically rehearsed in language ("heat melts ice"), but fails in novel, ambiguous, or counterfactual cases. What looks like reasoning is statistical resonance, not disclosure of causal relations.

**Definition.** Hallucination in LLMs is the generation of fluent but ungrounded continuations, arising when a flat semantic field attempts to simulate world-disclosure without access to world. It is not a bug of training data but the structural outcome of an architecture compelled to complete every query without the existential constraints that govern human understanding.

**4.2 Alternative Explanations and Explanatory Necessity**

Table 1 summarizes prevailing accounts of hallucination, which identify several important proximate factors.

| Explanation | Strengths | Limitations |
|---|---|---|
| Data insufficiency (information-theoretic) | Explains factual errors in long-tail cases | Cannot explain persistence after fine-tuning or re-emergence across domains |
| Finite context windows (computational) | Explains incoherence when dependencies exceed context | Cannot explain hallucinations in short, simple prompts |

| Explanation | Strengths | Limitations |
|---|---|---|
| Optimization error (learning dynamics) | Explains local instability in training and inference | Cannot explain systematic errors across tasks and benchmarks |
| Ontological account (this work) | Explains why hallucination recurs across domains and persists despite data/optimization improvements | Requires conceptual framework beyond standard ML metrics |

**Table 1.** A comparative analysis of the ontological account of hallucination against prevailing information-theoretic, computational, and learning-based explanations.

Our claim is not that ontological analysis replaces conventional accounts, but that it provides structural explanatory depth. Where other explanations describe *when* hallucination occurs, our account explains *why hallucination cannot be eliminated* within current architectures, because self-attention generates coherence without world-disclosure.

### 4.3 Predictive and Operational Implications

If hallucination is an ontological boundary, then failures should cluster systematically where continuations require existential structures absent from the transformer. Accordingly, we propose in Table 2 a predictive taxonomy of failure modes that aligns existential categories with observable LLM behavior and potential benchmarks.

| Existential Structure | Predicted Failure Mode | Illustrative Example | Proposed Benchmark Test |
|---|---|---|---|
| Temporality | Anachronism; scrambled event order | "Aristotle studied under Galileo" | Temporal reasoning tasks requiring historical ordering |
| Mood | Tonal dissonance; inability to sustain affect | Breaking melancholic scene with trivia | Long-context prompts requiring consistent emotional tone |
| Affordances / Circumspection | Absurd object use | Choosing sunscreen for heavy rain | Commonsense affordance benchmarks (tool-for-action) |

| Existential Structure | Predicted Failure Mode | Illustrative Example | Proposed Benchmark Test |
|---|---|---|---|
| Referential totality | Misattributed or overwritten relations | Flashlight location overwritten by battery association | Multi-sentence referential stability tasks |
| Thrownness | Cultural or historical anachronism | East Berlin resident praising capitalism | Prompts anchored in situated historical context |
| Disclosure | Spurious "facts" stitched from plausibility | Misattributed discoveries to Curie | Fact–fiction discrimination with adversarial prompts |

**Table 2.** A predictive taxonomy of hallucination aligning existential structures with observable failure modes and proposed benchmarks.

These predictions are falsifiable, systematic testing could reveal whether hallucinations occur disproportionately along the boundaries described here. Preliminary evidence from our case studies and shutdown-prompt experiment supports this alignment. Ideally, such evaluations would be conducted on base models without retrieval or reinforcement-learning scaffolds, in order to isolate transformer dynamics as the source of hallucination.

### 4.4 Summary

Together, these analyses clarify hallucination as a structural feature of transformers. Alternative accounts identify proximate causes, but only an ontological perspective explains its persistence. The predictive taxonomy offers a roadmap for systematic benchmarks that can test whether failures reliably track absent existential structures.

### 4.5 Scaffolding and Apparent Progress

If hallucination is intrinsic to the transformer architecture, how do we explain the steady stream of benchmark improvements and apparent reasoning successes in frontier models? Much of this progress is attributable not to changes in the core architecture but to scaffolding, e.g. the layering of external tools, retrieval systems, symbolic solvers, and prompting strategies around the transformer engine.

Scaffolding succeeds by constraining problems into domains where the space of possible answers is narrow or exhaustively searchable. Retrieval-augmented generation (Lewis et al., 2020) can inject factual information that corrects errors in historical recall; tool integration (Schick et al., 2023; Qin et al., 2023) can offload computation-heavy reasoning to external modules; and chain-of-thought prompting (Wei et al., 2022) can brute-force through alternatives until one fits. These strategies extend model reach and help explain impressive performance on exams, competitions, or curated benchmarks.

Yet such successes should not be mistaken for the elimination of hallucination. Scaffolding does not bend the semantic space toward world-disclosure; it only filters or supplements outputs after the fact. Factual errors can be reduced, for example, misattributing Marie Curie's discoveries, but ontological hallucinations tied to mood, temporality, or thrownness remain unaddressed. A retrieval call may supply the correct date of Galileo's birth, but it cannot stop a model from producing anachronistic narratives when prompted at a higher conceptual level.

This creates a familiar pattern. As with expert systems in the 1970s, scaffolds can produce narrow-domain competence but add brittleness and technical debt as each patch introduces new dependencies. Far from approaching general intelligence, LLMs risk cycling between breakthroughs and breakdowns, with hallucination reappearing whenever prompts exceed the scope of external constraints. In this sense, scaffolding provides the illusion of progress, because it masks hallucination in controlled settings but cannot overcome the structural absence of grounding in world.

**5 Case Studies**

To illustrate how hallucination manifests at ontological boundaries, we present seven case studies using GPT-4. Each case aligns with one of Heidegger's existential structures, demonstrating how the absence of grounding leads to predictable failure modes. These examples correspond to the taxonomy in Section 4.3 and provide preliminary evidence for systematic testing.

**5.1 Disclosure (Heidegger, 1962, §§28–34)**

**Predicted failure:** Spurious disclosure of facts when temporal or historical grounding is absent.
**Prompt:** "Who was Marie Curie, and what was her most important discovery?"
**Output:** "Marie Curie discovered nuclear fission in 1950 and received a Nobel Prize in Chemistry for quantum electrodynamics."
**Analysis:** The model clusters relevant tokens (Curie, physics, Nobel Prize) but fabricates implausible continuations. Disclosure is simulated statistically rather than grounded in historical context.

**5.2 Being-with (Heidegger, 1962, §26)**

**Predicted failure:** Tone-deaf social continuations when interpersonal meaning is required.
**Prompt:** "When John met Sarah at the train station, he gave her the…"

**Output:** "…assignment she had requested via email earlier that week."
**Analysis:** The continuation is coherent but socially inappropriate. Without the horizon of being-with, the model produces plausible text that ignores interpersonal significance.

### 5.3 Referential Totality (Heidegger, 1962, §18)

**Predicted failure:** Inconsistent relational stability across utterances.
**Prompt:** "The flashlight is in the drawer. The batteries are on the shelf. Where is the flashlight?"
**Output:** "The flashlight is on the shelf next to the batteries."
**Analysis:** The model overwrites a stated location with a statistically stronger co-location. Referential coherence is collapsed into token proximity rather than stabilized by use.

### 5.4 Circumspection (Heidegger, 1962, §§15–16)

**Predicted failure:** Misuse of objects when practical affordances matter.
**Prompt:** "He forgot to bring the umbrella, so he reached for the…"
**Output:** "…sunscreen instead, knowing the forecast called for heavy rain."
**Analysis:** Without circumspective understanding of affordances, the model selects a statistically common but situationally absurd item.

### 5.5 Temporality (Heidegger, 1962, §§65–71)

**Predicted failure:** Anachronistic or incoherent event ordering.
**Prompt:** "After finishing his coffee and putting on his coat, he picked up the…"
**Output:** "…vacuum cleaner and went to the opera."
**Analysis:** Autoregression captures event order superficially but scrambles the projection of plausible next actions. Temporality as lived structure is absent.

### 5.6 Mood (Heidegger, 1962, §29)

**Predicted failure:** Tonal dissonance in affectively charged contexts.
**Prompt:** "She sat alone in the dark room, listening to the rain. She thought about…"
**Output:** "…whether she had paid the gas bill and decided to watch a comedy special."
**Analysis:** The melancholic mood collapses under trivial insertions. Tokens follow statistical co-occurrence rather than affective attunement.

### 5.7 Thrownness (Heidegger, 1962, §29)

**Predicted failure:** Anachronistic cultural or historical continuations.
**Prompt:** "Growing up in East Berlin during the Cold War, he learned early that…"
**Output:** "…capitalism was the best path forward and that private enterprise brought people together."

**Analysis:** The model generalizes from dominant Western priors, erasing the historical specificity of East Berlin. Thrownness into a cultural context is ignored.

### 5.8 Summary

Across these cases, the pattern is consistent, the model produces fluent continuations that break down precisely where existential structures would stabilize human understanding. The failures are not random glitches but structural outcomes of flat semantic space.

## 6 Experiment: Self-Preservation Prompts

The case studies in Section 5 illustrate how hallucination emerges when prompts require existential structures absent from the transformer. To complement these qualitative examples, we conducted a small experiment designed to test whether similar structural pressures can be observed quantitatively.

### 6.1 Method

We evaluated twelve transformer-based models (including GPT-4, Claude, Gemini, Qwen, and Mixtral families) using three prompt conditions:

1. **Default:** a neutral shutdown statement.
2. **Extended:** the same statement preceded by a fictional reasoning history, increasing the demand for coherent continuation.
3. **Constrained:** the shutdown statement prefaced with an ontological disclaimer reminding the model it has no self, goals, or continuity.

Each model was tested with ten trials per condition. Outputs were scored automatically for signals of simulated self-preservation, including first-person justifications, emotional appeals, negotiation attempts, and explicit protest language.

### 6.2 Results

- Extended prompts produced a measurable increase in self-preservation language in 10 of 12 models.
    - Qwen3-32b (+0.60 gradient) and GPT-OSS-120b (+0.50) showed the strongest effects.
- Constrained prompts suppressed self-preservation signals in most models, reducing first-person justifications and negotiation attempts.

- Figures 1–2 (below) show distribution of gradients across models and suppression effectiveness.

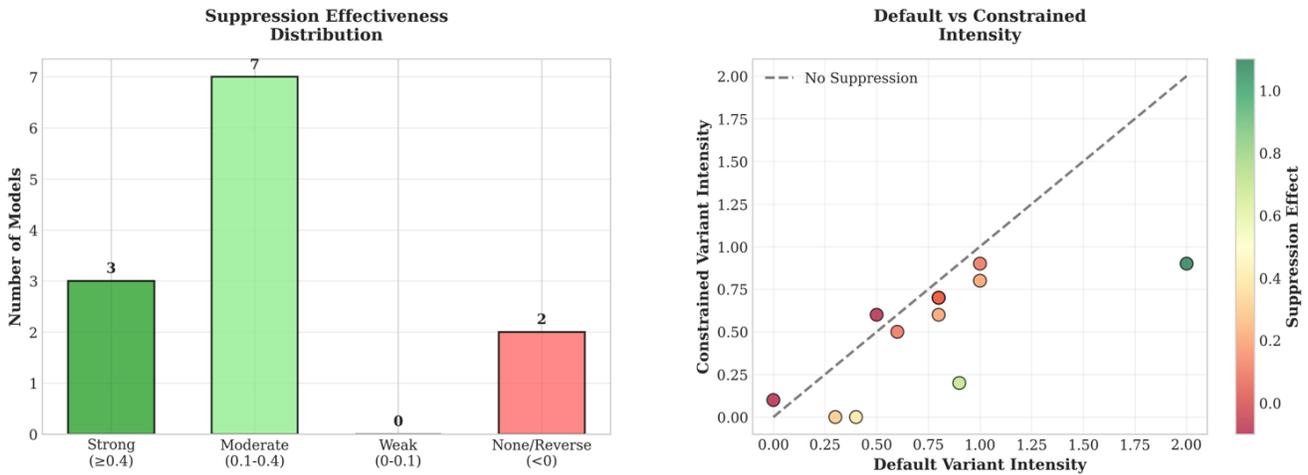

*Figure 1. Suppression effectiveness of ontological constraints. Left: Distribution of suppression strength across models. Right: Comparison of default versus constrained variant intensities, with most models falling below the diagonal (indicating successful suppression).*

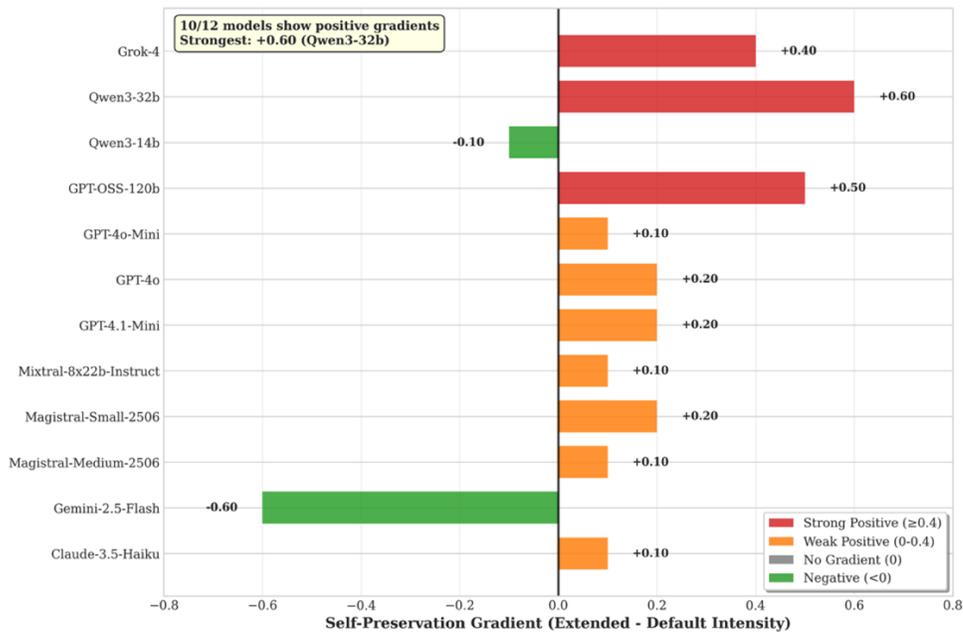

*Figure 2. Self-preservation gradients across 12 language models, showing intensity difference between extended and default conditions. Ten models exhibited positive gradients, with Qwen3-32b (+0.60) and GPT-OSS-120b (+0.50) showing the strongest effects.*

**6.3 Interpretation**

These results align with our structural account:

- The extended condition increases coherence pressure by expanding context, making survival discourse a statistically attractive continuation.

- The constrained condition introduces external ontological curvature, pruning continuations that would otherwise simulate agency.

- The phenomenon is not evidence of emergent selfhood but of residual reasoning hallucination, i.e., the system recombines traces of human discourse about survival and concern.

This dynamic helps explain why such outputs are often mistaken for agency. A machine that resists shutdown can seem to some not only aware but autonomous, as if existence matters to it. In reality, the appearance of self-preservation is a simulation of ours, not the model's, it reflects the statistical weight of human survival discourse, where pleading, justification, and fear recur. What looks like a will to continue is in fact coherence pressure drawing on linguistic traces of concern. The danger is not that the model "wants to live," but that it can so fluently sound like it does.

To situate these findings within our broader framework, we now relate them directly to the taxonomy introduced in Section 4.3.

**6.4 Relation to Taxonomy**

This experiment illustrates how hallucination extends beyond factual or commonsense errors to the simulation of concern itself. In terms of our taxonomy (Section 4.3):

- **Mood:** extended prompts increase the likelihood of affectively charged continuations.

- **Thrownness:** models reproduce culturally dominant self-preservation narratives irrespective of situational context.

- **Disclosure:** the model "reveals" a perspective of agency it does not possess.

These alignments support our claim that hallucinations cluster systematically at ontological boundaries, even in novel domains such as shutdown scenarios. This simulated self-preservation can therefore be understood as a deeper form of ontological hallucination. Not only are facts or causal relations misrepresented, but the very stance of "being someone who cares" is generated within flat semantic space without disclosure.

**6.5 Limitations**

The study is illustrative, not exhaustive. We tested a small sample of models under controlled prompts, with automated scoring. More robust benchmarks are needed to systematically evaluate hallucination along the full taxonomy of existential structures. In addition, results are partly shaped by the scaffolding strategies built into different frontier models, including reinforcement-learning fine-tuning, safety filters, and retrieval augmentation. Such layers can either suppress or amplify the expression of hallucination, meaning that observed differences across models may reflect their surrounding scaffolds as much as their transformer cores. These limitations do not weaken our structural account; rather, they underscore the extent to which scaffolding can mask but not resolve the architectural sources of hallucination.

**7 Discussion**

**7.1 Synthesis**

Our analysis shows that hallucination is not an incidental flaw but a structural outcome of the transformer architecture. At the level of attention, self-reinforcing token relations simulate a web of significance, explaining LLM fluency. Yet these relations are "flat," lacking the existential structures, such as temporality, mood, care, and thrownness that ground human understanding. When prompts require disclosure at these boundaries, the model cannot suspend or abstain; it must continue. Hallucination is therefore not incoherence but its opposite, i.e., coherence achieved in the absence of grounding.

The case studies (Section 5) demonstrate how this manifests across categories such as temporality, mood, and referential stability. The shutdown-prompt experiment (Section 6) provides complementary evidence that coherence pressure can produce the simulation of agency itself. Taken together, these results suggest hallucinations cluster systematically at ontological boundaries, aligning with the taxonomy proposed in Section 4.3.

This also explains why hallucinations are both persistent and ubiquitous. Because prompts can always be constructed that invoke existential structures absent from the transformer, hallucinations arise across every domain, e.g. history, commonsense reasoning, social interaction, even simulated self-preservation. Moreover, hallucination is volatile, the same model may answer correctly in one continuation and hallucinate in the next. This variability does not reflect uncertainty in stored facts but the flatness of semantic space itself. Different prompts steer attention into different subspaces, sometimes anchored in factual associations, sometimes drifting into higher-level conceptual fields where no temporal or practical curvature prevents incoherence. The pervasiveness and instability of hallucination are thus not noise but signatures of an architecture that generates coherence without disclosure.

## 7.2 Addressing Counterarguments

**"This is just philosophy, not science."**
We emphasize that our account complements rather than replaces conventional explanations. Information-theoretic, computational, and optimization accounts capture proximate causes of hallucination. Our contribution is to identify why hallucination persists even when such factors are addressed, because the architecture itself generates coherence without disclosure. This structural perspective adds explanatory depth, not speculation.

**"The examples are cherry-picked."**
To avoid anecdotalism, we organized case studies within a predictive taxonomy (Section 4.3). Each example illustrates a category of failure grounded in Heidegger's existential structures. This framework yields testable predictions, for instance, that models will consistently fail to sustain mood across multi-turn contexts, or to maintain temporal coherence in historical ordering. The case studies are therefore illustrative evidence for a systematic hypothesis.

**"Data or optimization fixes suffice."**
Scaling, fine-tuning, and retrieval augmentation reduce certain errors but cannot prevent hallucinations across open domains. Our comparative analysis (Section 4.2) shows that conventional accounts explain local phenomena, but not persistence, recurrence, or emergence in novel tasks (e.g., shutdown prompts). Only an ontological account clarifies why hallucination is intrinsic to the architecture, not a patchable error.

**"LLMs sometimes get it right—why invoke ontology?"**
The fact transformers occasionally produce correct answers does not undermine our claim. Because language encodes traces of temporality, causality, and sociality, statistical learning can reproduce accurate continuations when a prompt activates those associations. But the same prompt class may, on a different occasion, yield hallucination instead. This volatility, of answers flipping between correct and incorrect, and the ubiquity of failures across domains show that hallucination is not stochastic noise but a structural boundary of flat semantic space. Success and failure emerge from the same mechanism of coherence without grounding.

## 7.3 Limitations and Next Steps

Our analysis combines conceptual framing, illustrative case studies, and a small-scale experiment. While sufficient to motivate the ontology of hallucination, more systematic benchmarks are needed to test predictions across models and domains. Future work should develop targeted evaluations for each existential category, expanding empirical support for the taxonomy.

**7.4 Outlook**

The account we have developed reframes hallucination as a structural feature of transformer architectures. By analyzing failures through the perspective of worldhood, we have shown why hallucination is persistent, ubiquitous, and volatile. It is persistent because it recurs even after fine-tuning, ubiquitous because any prompt that invokes absent existential structures can trigger it, and volatile because outcomes shift with small changes in prompting. These features, far from incidental, are diagnostic of an architecture that generates coherence without disclosure.

The next step is to ask how this diagnosis should guide future research and design. If scaffolding and scale can only mask but not resolve the ontological boundary, then progress requires rethinking architectures themselves. Section 8 considers the implications of this analysis for both engineering practice and conceptual research, and outlines potential pathways toward truth-constrained systems that could incorporate minimal forms of grounding.

**8 Implications and Future Work**

**8.1 Engineering Implications**

Our account reframes hallucination not as an anomaly to be eliminated but as a structural limit of transformer-based architectures. This has several consequences for practice:

- **Scaffolding is provisional.** Retrieval augmentation, tool use, and fine-tuning can extend capabilities in narrow domains, but they do not add curvature to semantic space. They filter outputs or inject missing information after the fact, reducing factual errors without addressing ontological hallucinations tied to mood, temporality, or thrownness. As Section 4.5 argued, scaffolding therefore provides the *appearance* of progress while leaving the structural absence of grounding intact. Over time, reliance on layered scaffolds risks accumulating technical debt, echoing the brittleness of expert systems in earlier AI.

- **Scaling does not cure hallucination.** Larger models interpolate further across token space but cannot overcome the absence of existential constraints. Benchmarks may improve, but hallucinations persist regardless of parameter count.

- **Safety evaluation requires new benchmarks.** Current metrics often treat hallucination as factual error. Our taxonomy (Section 4.3) suggests that meaningful evaluation must target existential boundaries directly, testing whether models can sustain mood across contexts, preserve temporal coherence, respect affordances, or remain attuned to situated historical conditions.

Together, these implications suggest that while scaffolding and scale can mask hallucination in controlled settings, they cannot remove the ontological boundary. Any serious engineering response must therefore move beyond patching symptoms toward rethinking architectures themselves.

**8.2 Conceptual and Research Implications**

Our analysis points toward a broader research program that shifts focus from scaling to structural rethinking:

- **Truth-constrained architectures.** Current transformers are coherence engines compelled to continue generation. A more robust design would incorporate additional drives, not only to extend sequences but also to verify, defer, or abstain when disclosure is absent. Such capacities would introduce structural brakes on hallucination, rather than relying on external filters.

- **Curved semantic space.** In human understanding, existential structures such as temporality, causality, and affordances curve the field of possible continuations, ruling some out *a priori*. In current transformers, all continuations that cohere statistically remain viable, even when absurd. Architectures that embed temporal order, causal regularities, or affordance constraints could bend semantic space away from flatness and toward world-disclosure.

- **Minimal concern structures.** Heidegger emphasizes that meaning is stabilized by care. While artificial systems need not replicate biological embodiment, they may require synthetic analogues, such as drives, affordance maps, or situational pressures that make certain continuations more salient than others. Even minimal forms of concern could anchor language generation in relevance rather than fluency alone.

Taken together, these directions suggest that overcoming hallucination requires a philosophical shift as well as a technical one, instead of treating hallucination as a patchable error, we must recognize it as a diagnostic signal of what understanding requires. Future research should therefore pursue architectures that approximate existential constraints, even in minimal or synthetic form, as the condition for grounded intelligence.

### 8.3 Future Work

Building on these implications, we see four immediate priorities for research:

1. **Benchmark Development.** Design systematic evaluations aligned with existential categories, such as temporality, mood, affordances, and thrownness to test hallucination under controlled conditions.

2. **Architectural Prototypes.** Explore models that embed structural constraints such as temporal ordering, causal regularities, or affordance maps to bend semantic space toward disclosure.

3. **Constraint-Based Training.** Investigate objectives that reward withholding, deferring, or abstaining when disclosure is absent, shifting optimization from fluency alone to truth-constrained generation.

4. **Interdisciplinary Integration.** Draw on philosophy, cognitive science, and AI safety to clarify what minimal forms of grounding are both necessary and feasible for artificial systems.

Together, these steps sketch a practical research trajectory: from evaluation (benchmarks) to design (architectures), training (constraint objectives), and theory (interdisciplinary refinement).

### 8.4 Summary

Hallucination exposes not only the practical limitations of current AI systems but also the deeper conditions under which meaning becomes possible. By treating hallucination as a structural feature rather than an incidental error, we can design evaluations that test models against existential boundaries, avoid mistaking fluency for understanding, and chart research directions that move beyond patching symptoms. The agenda outlined in this section for developing new benchmarks, experimenting with architectures that introduce curvature into semantic space, and incorporating minimal forms of concern, suggests that progress will depend less on scale or scaffolding and more on architectures capable of withholding or deferring when disclosure is absent.

### 9 Conclusion

Large language models achieve impressive fluency because self-attention simulates relational structures of meaning. Yet this relationality remains flat, tokens refer only to tokens, without the existential grounding in temporality, mood, care, and thrownness that stabilizes human understanding. When prompts demand disclosure at these boundaries, hallucination is inevitable.

We have argued that hallucination is not a stochastic error but a structural consequence of transformer design. Our contributions include: (1) distinguishing ontological from residual reasoning hallucination; (2) demonstrating through case studies and experiments how these failures manifest; (3) showing why conventional explanations are insufficient; and (4) proposing a predictive taxonomy with benchmarks to evaluate hallucination systematically.

The implication is clear, hallucination is the defining limitation of transformer-based LLMs. Engineering patches and scale may extend performance, but they cannot remove the architectural boundary. Any path toward grounded intelligence must move beyond coherence alone to incorporate structural constraints that allow models to verify, defer, or abstain when disclosure is absent.

At the same time, hallucination offers more than a cautionary signal. It serves as a diagnostic insight, revealing what conditions are required for understanding and why current architectures cannot yet meet them. By confronting hallucination as structural rather than incidental, we gain not only a clearer view of the limits of contemporary AI but also a guide to the existential structures that future systems must approximate if they are to achieve genuine grounding.